# Eye-GUIDE (Eye-Gaze User Interface Design) Messaging for Physically-Impaired People

Rommel Anacan[1], James Greggory Alcayde[2], Retchel Antegra[3] and Leah Luna[4]

[1]Electronics Engineering Department Technological Institute of the Philippines Manila, Philippines

rommel.anacan@ieee.org
james_jhemz02@yahoo.com.ph
aerial_damn@yahoo.com
savannah_ley@yahoo.com

## ABSTRACT

*Eye-GUIDE is an assistive communication tool designed for the paralyzed or physically impaired people who were unable to move parts of their bodies especially people whose communications are limited only to eye movements. The prototype consists of a camera and a computer. Camera captures images then it will be send to the computer, where the computer will be the one to interpret the data. Thus, Eye-GUIDE focuses on camera-based gaze tracking. The proponent designed the prototype to perform simple tasks and provides graphical user interface in order the paralyzed or physically impaired person can easily use it.*

## 1. INTRODUCTION

Eye tracking system is a communication and control system for people with physical disabilities. It has been studied by many for the past years because of its potential applications in different fields of eye-movement analyses especially in the field of medicine such as Human-Computer Interface (HCI), Eye Disease Diagnosis, and other related disciplines. Eye tracking is a technique that acquires and analyzes the eye movements and use it to determine where user's attention is focused through his eyes which can be used to manipulate the cursor position relative to the position of the user's eye on the screen. Eye movement tracking is a technology that can be used in navigating the computer screen without the need for mouse or keyboard input, which can be benefecial for certain populations of users such as disabled individuals[13].

Tracking eye movements can help understand visual and display-based information processing and the factors that may impact upon the usability of system interfaces [17].

Eye tracking has been growing for years and new approaches on interfacing have emerged. One of these applications is the manipulation of mouse cursor through eyes, eliminating the need for a computer mouse which is primarily for disabled people [13].

The life of a disabled individual has been limited due to his/her disability. People who have lost their hands and so are the people who were born to have defects in their hands have become limited from holding capabilities, depriving them from a lot of things. Eye tracking technology has evolved over years and this technology can alleviate the condition of these people [9].

The paper aims to provide solutions to present problems and needs, to suffice or provide basic and extensive knowledge regarding eye tracking, to improve health condition of the people, to enrich research methods for eye tracking, to present importance for the respondents of this study and to somehow support government thrusts. Inasmuch as the United Nations classified access to the Internet as a Human Right, people, who became limited due to incapacity of holding/handing things as an effect of their conditions, was not able to enjoy this right. By developing a system such as mouse controlled by eyes can bridge this gap which therefore contributes in supporting government thrusts.

The prototype of the thesis is subdivided into two parts—hardware and software implementation. Eye-GUIDE focuses on video-based gaze tracking, which consist Hannah camera that records the eye of the user and a computer. The webcam can be used with or without infrared (IR) illumination. It is mounted into an eyewear. The main reason why the webcam is mounted is to avoid misalignment of the camera lens that will result to lost-focus onto the human eye pupil. The camera capture images, which is transferred to a computer.

For the software implementation the proponents used three different softwares for eye gazing, eye clicking and messaging system. The software for eye gazing will extract some eye features from the image captured by the camera, such as pupil centre or iris centre. The direction of gaze estimation is achieved from the pictures acquired by the camera. This is analyzed with the aid of intense color contrast of the white sclera and black iris which helped in iris detection but due to the presence of eyelid, accuracy is high only for portions that are not covered by the eyelid (in instances of blinking) which is only for horizontal portions. Calibrations keep track of eye features during the process. The result of this process tells the accuracy of the eye to locate circle in the different spot on the screen. The result may not be good enough for excellent and stable eye tracking due to noise that contaminate gaze data acquired and detection of fixation may help to deal with this. Upon consideration of this, saccades and fixations are factors to be noted which are both related to eye movements. For clicking purposes, the Eye GUIDE Clicker software was introduced. This tool allows user to click templates, keyboard letters and supporting buttons on the Eye GUIDE Messenger provided by this study by hovering the cursor several seconds over the desired button or icon. Finally, software for this messaging system allows the user to select from the provided templates, what the user wanted to say to his guardian or to his physician or nurse. These templates are to be translated by the system from text to voice. The messaging system also provided a screen keyboard to let the user to type his taught that is not included on the provided templates. Lastly, the system has an alarm feature to call for attention.

## 2. METHODOLOGY

### 2.1 Research Design

The research design provides the glue that holds the research project together. A design is used to structure the research, to show how all of the major parts of this thesis project -- the samples or groups, measures, treatments or programs, and methods of assignment -- work together to try to address the central research questions. For this research, proponents have used experimental design.

To be able to make this thesis possible, the proponents made plan of action on how this thesis will be conducted. First step done is the basic and fundamental foundation of every research. The proponents conducted data gathering through reading of several books, journals, magazines, published and unpublished study to acquire information regarding cursor manipulation and eye tracking system. With the use of these materials, it has been concluded that there are several ways in making an eye controlled mouse. After reading these materials we have taken several ideas that were basically paraphrased.

One approach is through the use of video-based tracking [15]. For this kind, using eye-tracking tool, such as camera, the system was able to notice precisely where the user's eye is staring on the screen. The binocular eye-tracker has been configured to send x and y coordinates that is used to determine eye position with respect to the screen [10]. Software on the PC uses this information and sends it to Windows as a mouse signal, and the x, y coordinates determine the cursor position to the screen. The selection (the emulation of pressing the button on a mouse) is achieved by the software detecting an unnatural event. This unnatural event refers to the time when the person winks (different from natural eye blink and the eye tracker won't treat blinking as a valid signal) or when the eye is held still for half a second. The use of a binocular system allows the closure of the right and left eyes to emulate pushing the right and left mouse buttons respectively [17].

Electrodes are also used for eye tracking. This method acquires EOG signal from the moving retina of the eye which is proved to induce potential to the electrodes. Through this signal, vertical and horizontal movements of the eye can be used for cursor manipulation and eye blink for clicking. There are also studies that able to discriminate eye winks to eye blinks to avoid unintended clicks [18].

For this particular study, proponents used camera-based eye tracking. It is another approach for eye tracking that is dependent on the light reflecting capabilities of the eye. Several techniques can be used under this approach. One is pupil tracking that this study has implemented for the prototype. Pupil of the eye can reflect lights which can be used for tracking the movements of the eye [16].

Upon detection of the eye movement, signals are processed, amplified and digitized to be used for cursor manipulation [18]. This stage prepares the signal for analyses for proper control of the mouse cursor. It involves noise reduction and signal identification.

The output of the second stage is inputted to the software that is necessarily needed to interpret the activity of the eye made by the user. For interpreting these data, there has been different software available. NI LabView, VHDL, HPVEE, math lab, and flow code are some of them [2]. Nl LabView was used over another popular data acquisition tool, due to its more extensive graphical capabilities along with its availability [1]. LabVIEW is necessary to convert the signal obtained by the EOG into interpretable data for directional discrimination. Moreover, a graphical display will be implemented in LabVIEW to simulate the movement of an icon on the computer screen [15]. For this study, C# Visual Studio 8 is used for eye tracking software and Visual Basic for provided messaging interface.

After the final stage, when the research is made, the design is tested and evaluated for its reliability and accuracy to perform the objectives of the study. Mouse movements can be used to infer a user's intent and focus while browsing a website. By using mouse movements in usability testing, researchers can determine if users' expectations are met. This tool can be especially beneficial in conjunction with other techniques used in usability testing, such as think aloud procedures, as this information can lead to a better mouse movement model.

## 2.3 Research Setting

The design and development was literally conducted in our school, Technological Institute of the Philippines, as the basic requirement for the subject Project Study. Graduating students must take up this subject and should propose and complete a certain research, technical research for that matter that is highly involving a great advance of technology. The thesis must also address a certain problem that will be probably solving that problem though the prototype is made in one of the proponent's place to be able to work promptly and safely. The testing will be done also in the school to present the thesis to the panels.

## 2.4 Respondents of the Study

The prime objective of the thesis is to be able to produce a prototype that will enable to control the cursor of the mouse through transduction of eye movements. With this concern, it was aimed to help people with disabilities especially those who have incapacitated hands and people who lost their hands due to different circumstances or hereditary cases. Moreover, the study will also help reduce cases of ergonomic hazards like carpal tunnel syndrome as encountered by people who frequently use computer mouse or use this for a long duration of hours. For these cases, the respondents of the study would be people with or without disabilities since everyone can be categorized as people who might have risk on ergonomic hazard.

## 2.5 Data Gathering Procedure and Instruments Used

This portion contains a simple discussion as to how the data were being gathered and analysed in order to fully understand the methods being used in this particular thesis. It contains two parts. First part is the clerical and the second one contains all the mechanical devices being used.

**2.5.1 Clerical Tools**

The testing for this thesis can be done for those people with disabilities who, by their conditions, are refrained from handling traditional mouse. On the contrary, the objectives of the study is not limited for such people because it can also be used by people with normal conditions to test the effectiveness of this study and to assess the workability of the design for the convenience of the users. The desired output should be met in order to satisfy the expectations of the respondents.

The study made a Graphical User Interface (GUI) named Eye GUIDE Messenger for the selection of icons pertaining to the type of action involved. The GUI has a screen keyboard so that the respondents can type in anything he or she wants to tell. It also has ten commands on the right side of the computer screen which are converted to audible sound upon clicking. The respondents will just have to focus the mouse cursor on the command he or she likes for several seconds. It will be then automatically clicked.

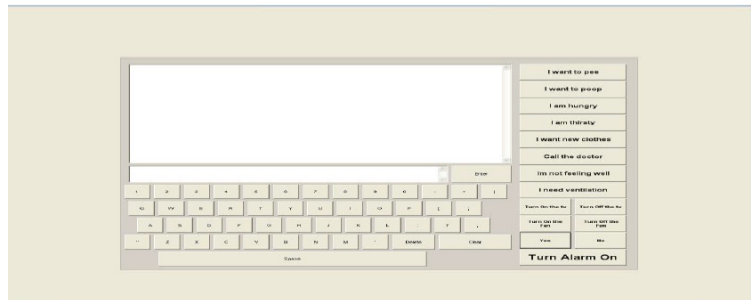

**Figure 1.** Eye GUIDE Messenger

**2.5.2 Mechanical Devices**

**2.5.2.1 Camera for Eye Tracker**

The proposed system utilizes only one web camera to acquire user's eye images. To eliminate influences due to illumination changes, the camera which includes IR emitting diode is used because this helps compensate for such effects. User navigates on the screen while the camera keeps on tracking. The camera is attached to the head of patient by means of the glasses frame. The interface shows an input live video of the user's eye region in the first frame, and the eye tracking result in the current frame.

**2.5.3 Computer for the Graphical User Interface (GUI)**

The Graphical User Interface (GUI) can be installed and run in either a laptop or a desktop.

## 2.6 Flowchart of the Prototype

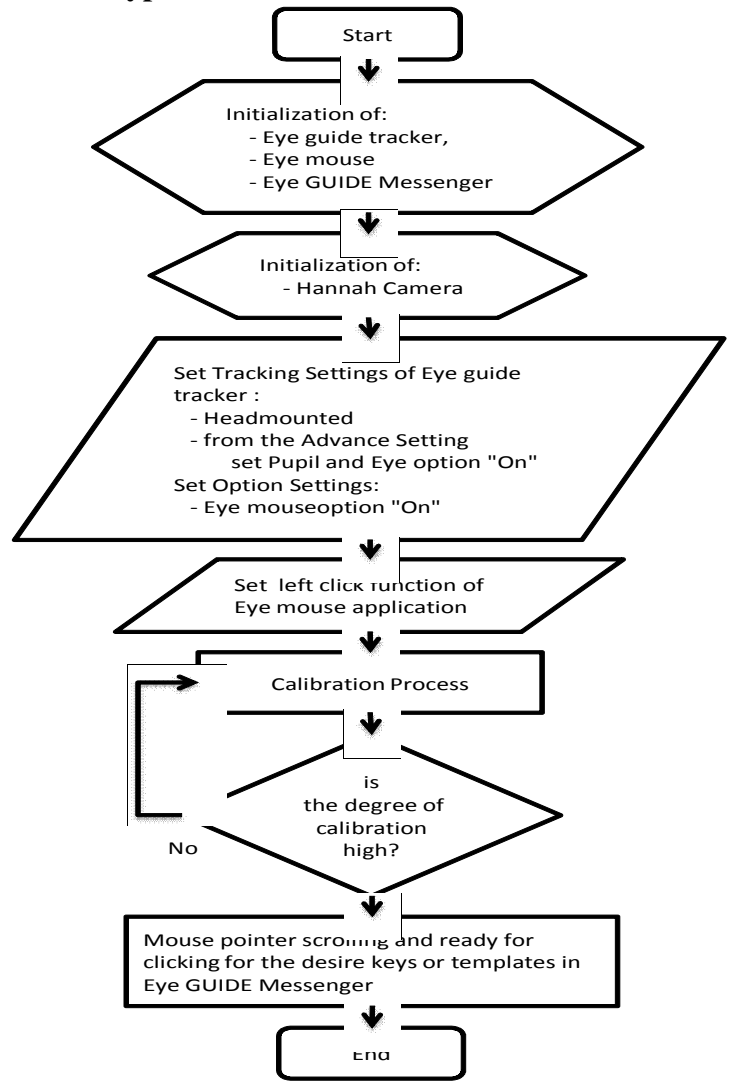

**Figure 2.** System Flow Chart

## 2.7 Technical Design

Eye-GUIDE (Eye-Gazed User Interface Design) is a tool that involves three applications which are Eye GUIDE Messenger, Eye GUIDE tracker and Eye mouse. Initially, Hannah camera will be initialized as we open the Eye guide tracker application. When the camera has been initialized, the system should be set the following parameters; a) Set Tracking Settings as "Head mounted" and from the Advance Setting set Pupil and Eye option "On". Then, in the Option Settings, set Eye Mouse option "On". When all parameters have been set correctly, the system is ready for calibration. Take note the Eye-Guide Tracker tool will not start until calibration process was not finish. Lastly, make sure that the Eye mouse is set to left click function since the provided graphical interface only need to be click once.

When the user got a failing calibration rate he/she must recalibrate the system to achieve high accuracy gazing of the mouse. When the user got a high calibration rate he/she may now use the tool. Mouse pointer scrolling clicking is ready to use for the selecting the desired keys or templates in Eye GUIDE Messenger.

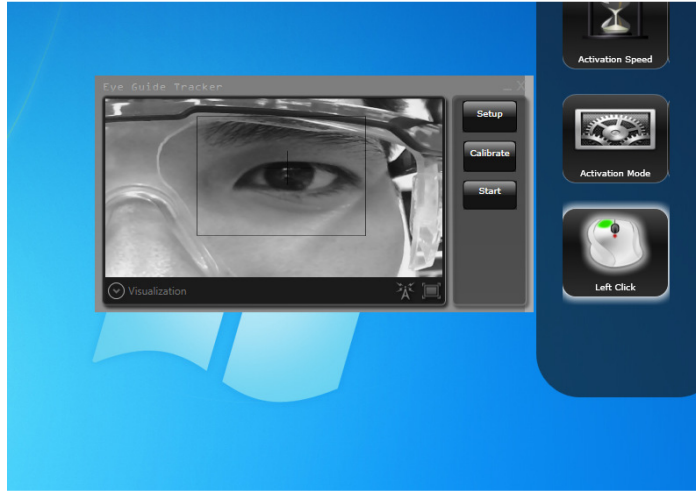

**Figure 3.** Eye GUIDE Tracker (centre image) and Eye GUIDE Clicker (right image)

## 2.8 Statistical Treatment

The eye tracking system design has no correlation with respect to the distance and the angle of the face of the user from the computer used because the main camera that is used to track the pupil of the eye of the user is attached to the eyewear created and since it is worn by the user, it is considered to have a fixed position all throughout the testing of the device, with respect to the user's eye. Moreover, the winks and blinks done by the user has no significant effect on the reliability and effectiveness of the design since these actions have minimal effects on pupil tracking since these are done very quick in range of nanosecond that the system can be able to track again the pupil once the eye is open as long as there attained a good calibration results.

All the data of the thesis were taken only from a specific respondent because the system used was considered as new technology that not everybody can easily perform the functions of the system. The respondent of the thesis is the one who has better capability of controlling his eye which is tracked during the whole process.

| Distance | Time to Reach an Icon and Click it (in seconds) | | | |
|---|---|---|---|---|
| From the Computer | First Trial | Second Trial | Third Trial | Average Time |
| 12 in | 1.5 | 2.2 | 1.07 | 1.59 |
| 24 in | 2.44 | 1.98 | 2.1 | 2.17 |
| 36 in | 3.13 | 1.22 | 2.28 | 2.21 |

**Table 1.** Data and Results for Testing First Hypothesis

All the results for this hypothesis testing are taken into average so that the precision of the performance of the design is examined. There are three trials for this test. It can be seen that there were different results for every trial since there were factors to be considered. These factors were the illumination of the surrounding, sensitivity of the camera, the degree of exactness from the calibration of the device, the size of the icon to be click and the place of the mouse before the timing is started.

The first distance considered is 12 inches away from the computer screen. On the first trial, the time it takes to click the icon is 1.5 seconds, 2.2 seconds on the second and 1.07 seconds on the third with an average of 1.59 seconds. There is an easy time targeting the icon since the closer the eye is to the computer, the greater the movements that the camera will track. Therefore, the mouse is optimally controlled.

Doubling the distance, 24 inches away from the computer screen, is examined next. It takes 2.44 seconds for the icon to be clicked on the first trial, 1.99 seconds on the second and 2.1 seconds on the third with an average of 2.17 seconds. It can be seen that the first distance considered is better in terms of time response. Though the difference in the time response with this distance is small, the eye movement is better tracked due to more movements being tracked, making it easier to manipulate the cursor but the icons are smaller at this distance.

For the last distance to be experimented, increasing the distance by 100% or 36 inches away from the computer screen, the results became worse. 3.13 seconds is the time it takes for the icon to be clicked on the first trial, 1.22 seconds on the second and 2.28 seconds on the third with an average of 2.21 seconds. Though the device still working on this distance, it became harder for the respondent to click the icon since the icon became smaller at this distance which refrained from tracking more movements of the eye and eventually degrade the performance of the system at small difference in time response.

| Angle of the Face | Performance Data | | |
| --- | --- | --- | --- |
| | Pupil Detection (Y/N) | Cursor Manipulation (Y/N) | Time to Reach and Click an Icon (in seconds) |
| 30$^o$ | Y | Y | 2.3 |
| 45$^o$ | Y | Y | 3.8 |
| 90$^o$ | Y | Y | 1.5 |

**Table 2.** Data and Results for Testing Second Hypothesis

For this hypothesis testing, performance is evaluated through the ability of the device to still detect the pupil of the respondent and manipulate the cursor, and the time response in clicking the icon. It can be seen that even the face is tilted at certain angles; the tracking ability is still not compromised due to eyewear being worn and stabilized before calibration. The time it takes to click the icon with the face tilted at 30$^o$ is 2.3 seconds, 3.8 seconds for the face being tilted at 45$^o$ and lastly 1.5 seconds with the face not being tilted at 90$^o$. The factors to be considered here are illumination of the surrounding, sensitivity of the camera, the degree of exactness from the calibration of the device, the size of the icon to be click and the place of the mouse before the timing is started.

## 3. TESTING RESULTS

### 3.1 Camera Testing

It has been mentioned that the study uses a Hannah camera. The study successfully demonstrated that the system can track the user's eye features under various environments with complex background, such as office, darkness environment and indoor environment. As results, there had been differences on every environment due to illumination, sensitivity and system calibration before mouse cursor manipulation. Despite environment change, based on illumination recognition, the system can still obtain the visual tracking of the eye pupil features.

### 3.2 C# Visual Studio 8 Testing

The implementation of C# Visual Studio 8 programs worked well. The software for eye gazing will extract some eye features from the image captured by the camera, such as pupil centre or iris centre. The direction of gaze estimation is achieved from the pictures acquired by the camera. This is analyzed with the aid of intense color contrast of the white sclera and black iris which helped in iris detection but due to the presence of eyelid, accuracy is high only for portions that are not covered by the eyelid (in instances of blinking) which is only for horizontal portions. Calibrations keep track of eye features during the process. The result of this process tells the accuracy of the eye to locate circle in the different spot on the screen. The result may not be good enough for excellent and stable eye tracking due to noise that contaminate gaze data acquired and detection of fixation may help to deal with this. Upon consideration of this, saccades and fixations are factors to be noted which are both related to eye movements. For clicking purposes, the Eye GUIDE Clicker software was introduced. This tool allows user to click templates, keyboard letters and supporting buttons on the Eye GUIDE Messenger provided by this study by hovering the cursor several seconds over the desired button or icon. Finally, software for this messaging system allows the user to select from the provided templates, what the user wanted to say to his guardian or to his physician or nurse. These templates are to be translated by the system from text to voice. The messaging system also provided a screen keyboard to let the user to type his taught that is not included on the provided templates. Lastly, the system has an alarm feature to call for attention.

## 4. SUMMARY OF FINDINGS

From the test of the device, the effectiveness of the project is evaluated and the reliability of the design is assessed. From the evaluation of the effectiveness of the device, certain factors were analyzed and were found out. The factors that affect the workability and efficiency of the device were found to be, first is the illumination of the surrounding. Illumination is the prime factor for eye tracking system using a camera. Since the camera track the light reflection from the pupil, the eye to be tracked being illuminated by a greater intensity degrades the tracking ability of the camera which hinders good performance for the system.

Second factor is the sensitivity of tracking mechanism of the camera. Since the camera is greatly dependent to the illumination, the sensitivity of the camera is affected resulting to unstable operations. The solution for this is manual adjustment of the sensitivity which is rendered by the software provided by this study and calibration prior to cursor manipulation.

Inasmuch as the device is evaluated through the time response, which is acquired from recording the time it takes for the respondent to click the icon, the size of icon to be clicked is one factor for this and the distance of the respondent to the computer screen. It is easier to click a small icon than a large icon but large icons are easier to focus. The best time response is best acquired at closer proximity to the screen as seen from the results of first hypothesis testing.

# 5. CONCLUSION AND RECOMMENDATION

The mouse pointer controlled by the eye has been successfully simulated and was able to test the effectiveness of the prototype as a basic interfacing tool for Eye GUIDE Messenger developed. The study has successfully demonstrated that the prototype can track user eye features and detect the pupil under various environments with complex background having clutters. These two variables can also be acquired with various distances and various angle of the face of the user being tracked. The system had been calibrated enough to ensure that the user was able to move the cursor through his eye so that the system can track its movement. The study showed successful operations with proper calibration and enough practice for the device usage.

The Eye GUIDE showed that it has a potential to be used as generalized user interface in many applications. For further improvements, this study suggests creating a better eye tracker system that has higher sensitivity to attain excellent calibration results under various illumination effects. Moreover, the calibration step can be made to operate automatically upon installation of the tool so that the user will no longer entail manual adjustments to deal with different illumination of the surroundings to produce a more user friendly eye tracking system.

# ACKNOWLEDGMENT

This study will never be possible without the help and assistance of few people that had guided us in our way in probing the right knowledge for this design and achieving the best paper for this study. For this, we would like to express our deepest indebtedness to following with the highest deference.

Our sincerest gratitude, first, to God for guiding us and sufficing all the knowledge and understanding on this study and most especially for providing us the people who helped from concept development to manufacturing to final defence of this paper.

We will be forever grateful to our families for unconditional love, care and support, for unquestioning financial assistance and for believing in our abilities. Also, to our friends, thank you all for backing up in every endeavour we make and for being there in rocks and success.

# AUTHORS

**Rommel M. Anacan** was born in Manila, Philippines in 1986. He received the B.S. degree in electronics engineering from Pamantasan ng Lungsod ng Maynila, Manila, Philippines in 2008 and the M.S. in electronics engineering major in microelectronics from Mapua Institute of Technology, Manila, Philippines in 2011 . He was a recipient of DOST-Scholarship for his Master's program. Currently, he is pursuing his Ph.D in electronics engineering at the same university.

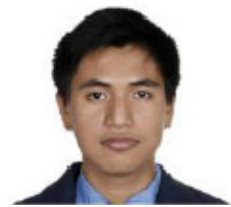

**James Gregorry Z. Alcayde** was born in Bulacan, Philippines in 1988. He received the B.S. degree in electronics engineering from Technological Institute of the Philippines-Manila, Manila, Philippines in 2012.

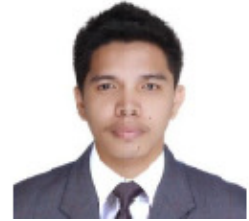

**Retchel C. Antegra** was born in Cavite, Philippines in 1989. She received the B.S. degree in electronics engineering from Technological Institute of the Philippines-Manila, Manila, Philippines in 2012.

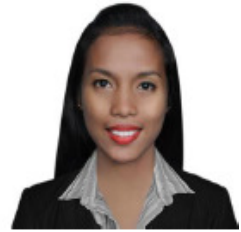

**Leah J. Luna** was born in Quezon City, Philippines in 1988. She received the B.S. degree in electronics engineering from Technological Institute of the Philippines-Manila, Manila, Philippines in 2012.

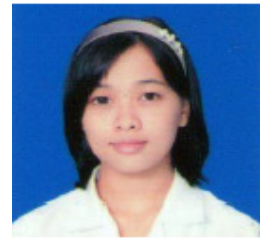